\documentclass[preprint,preprintnumbers, prd, floatfix, superscriptaddress,nofootinbib] {revtex4-1}
\usepackage{epsfig}
\usepackage{subfigure}
\usepackage{dcolumn}
\usepackage{bm}
\usepackage[usenames ,dvipsnames]{xcolor}
\usepackage{slashed}
\usepackage{graphicx,color}
\begin{document}
\title{New $X_{0,1}(2900)$-like exotic states in $b$-baryon decays}

\author{Yu-Kuo Hsiao}
\email{Corresponding author: yukuohsiao@gmail.com}
\affiliation{School of Physics and Information Engineering, Shanxi Normal University, Linfen 041004, China}

\author{Yao Yu}
\email{Corresponding author: yuyao@cqupt.edu.cn}
\affiliation{Chongqing University of Posts \& Telecommunications, Chongqing 400065, China}

\date{\today}

\begin{abstract}
In the $B^+\to D^+ D^-K^+$ decay,
LHCb has reported the observation of
the open-charm exotic states $X_{0,1}^0\equiv X_{0,1}(2900)^0$
with four different quark flavors~$(ud\bar s \bar c)$,
where the subscripts (0,1) denote the spins.
To confirm the discovery,
we propose $\Lambda_b\to \Sigma_c^{0(++)} X_{0,1}^{\prime\,0(--)}$
in the final state interaction,
where $X_{0,1}^{\prime\,0(--)}$ with
$s u\bar d \bar c$ ($ds\bar u \bar c$)
are the new $X_{0,1}$-like exotic states. More specifically,
$\Lambda_c^+D_s^-$ in $\Lambda_b\to \Lambda_c^+D_s^-$
are transformed as
$\Sigma_c^{0(++)} X_{0,1}^{\prime\,0(--)}$, by exchanging $\pi^{+(-)}$.
As the order of magnitude estimates, we calculate
${\cal B}(\Lambda_b\to \Sigma_c^{0(++)} X_{0,1}^{\prime\,0(--)})
=(2.3\pm 0.6,4.3\pm 0.8^{+3.3}_{-2.5})\times 10^{-4}$.
In addition, we estimate other $b$-baryon decays with the $X_{0,1}$-like states,
such as ${\cal B}(\Xi_{b}^{0(-)}\to \Xi_{c}^{0(+)}(2645)X_{0}^{\prime\,0(--)},
\Lambda_{b}\to \Xi_{c}^{\prime\,0}X_{0,1}^{0})\sim 10^{-5}$.
While one needs $\Lambda_b\to \Sigma_c^{0(++)}M_c M$
to observe  $X_{0,1}^{\prime\,0(--)}\to M_c M$ with
$M_c M=D_s^-\pi^+$, $\bar D^0 \bar K^0$ $(D_s^-\pi^-$, $D^-K^-)$,
${\cal B}(\Lambda_b\to \Sigma_c^{0(++)} X_{0,1}^{\prime\,0(--)},
X_{0,1}^{\prime\,0(--)}\to M_c M)\sim 10^{-4}$ are accessible to the LHCb experiment.
\end{abstract}

\maketitle
\section{introduction}
The LHCb Collaboration has recently observed $X_{0,1}^0\equiv X_{0,1}(2900)^0$
from the $B^+\to D^+D^-K^+$ decay~\cite{Aaij:2020hon,Aaij:2020ypa},
where the subscript 0(1) denotes the spin.
With the resonant strong decays $X_{0,1}^0\to D^- K^+$
in the $D^- K^+$ invariant mass spectrum,
one concludes that $X_{0,1}^0$ are the first open-charm exotic states,
which consist of four quarks with different flavors $\bar c d \bar s u$.
Explicitly,
the masses and decay widths of $X_{0,1}^0$ are
given by~\cite{Aaij:2020hon,Aaij:2020ypa}
\begin{eqnarray}\label{MandW}
(m_{X_0^0},m_{X_1^0})&=&(2.866\pm 0.007\pm 0.002,2.904\pm 0.005\pm 0.001)~\text{GeV}\,,\nonumber\\
(\Gamma_{X_0^0},\Gamma_{X_1^0})&=&(57\pm 12\pm 4,110\pm 11\pm 4)~\text{MeV}\,,
\end{eqnarray}
where $m_{X_0^0}\simeq m_{X_1^0}$ and $\Gamma_{X_0^0}\simeq \Gamma_{X_1^0}/2$.
From the strong decays $X(5568)^\pm\to B_s^0\pi^\pm$,
the ``open-beauty'' exotic state $X(5568)$
with four different quark flavors $\bar b s \bar d u$ $(\bar b s \bar u d)$
was once reported to be observed by D0 Collaboration~\cite{D0:2016mwd}.
Unfortunately, the discovery was not confirmed by LHCb and other experiments~\cite{Aaij:2016iev,
Sirunyan:2017ofq,Aaltonen:2017voc,Aaboud:2018hgx}.
Since it is possible that
$X_{0,1}^0$ can be further examined as the tetraquarks,
which definitely improves our knowledge of QCD and the quark model,
one needs to provide the different decays to confirm the discovery.

Theoretical attempts have been given
to understand the open-charm exotic states~\cite{Molina:2010tx,Karliner:2020vsi,He:2020jna}.
Considering the light quark $q=(u,d,s)$ as a triplet (3) under
the $SU(3)$ flavor symmetry ($SU(3)_f$),  the open-charm exotic states
$q_1 q_2 \bar q_3\bar c$ are in the irreducible forms of
$(3\times 3\times \bar 3)\bar c=(3+3+\bar 6+15)\bar c$,
where $\bar 6$ and 15 are able to include three different quark flavors.
Consequently, one classifies $X_0^0$ and $X_1^0$ into $\bar 6$ and $15$,
with $X_0^0=(ud-du)\bar s\bar c$ and $X_1^0=(ud+du)\bar s\bar c$,
respectively~\cite{He:2020jna}.
Moreover, $X_{0(1)}^0$ is assigned the quantum numbers $J^P=0^+(1^-)$,
which is based on the first radially (orbitally) excited state $2S$ ($1P$)
of the two-body Coulomb and chromomagnetic interaction models.
In this classification,
there can be $X_{0,1}$-like exotic states,
\begin{eqnarray}\label{su3}
X^{\prime\,--}_{0,1}=(ds\mp sd)\bar u\bar c\,,\;
X^{\prime\,0}_{0,1}=(su\mp us)\bar d\bar c\,,\;
Y^{\prime\,-}_{0,1}=(sn-ns)\bar n\bar c\,,
\end{eqnarray}
with $J^P=(0^+,1^-)$, $n=u$, $d$, and $m_{X_{0(1)}^\prime}\simeq m_{X_{0(1)}}$.
In addition, $X_{0(1)},X_{0(1)}^\prime\to DM$ are related with
the same strong coupling constant. Therefore,
the observation of $X'_{0,1}$ can be regarded to confirm
the discovery of $X_{0,1}$.

As the discovery channel,
$B^+\to D^+X_{0(1)}^0, X_{0(1)}^0\to D^-K^+$ has been
interpreted to proceed through the final state interaction~\cite{Burns:2020epm,
Burns:2020xne,Chen:2020eyu,Liu:2020orv}.
Following the $B^+\to D_s^{*+} \bar D^0$ weak decay,
$D_s^{*+} \bar D^0$ in the rescattering effect
are transformed as $D^+ X_{0,1}^0$, which is by exchanging $K^0$.
Moreover, the same mechanism
has been applied to the production of  the hidden-charm (strange) pentaquark
${\cal P}_{c(s)}$ with $c\bar c uud$ ($c\bar c uds$)
in $\Lambda_b\to J/\Psi {\cal P}_{c(s)}$~\cite{Wu:2021dmq,
Liu:2015fea,Aaij:2020gdg,Aaij:2015tga}.
Therefore, it is reasonable to consider the $b$-baryon decays
with the new $X_{0,1}^0$-like exotic states in the final state interaction.
Specifically,
we propose $\Lambda_b\to \Sigma_c^{0(++)} X_{0,1}^{\prime\, 0(--)}$
as depicted in Fig.~\ref{triangle},
with $X_{0,1}^{\prime\,0}$ and $X_{0,1}^{\prime\,--}$ in Eq.~(\ref{su3}).

To estimate $\Lambda_b\to \Sigma_c^{0(++)} X_{0,1}^{\prime\, 0(--)}$,
the couplings in the triangle loop should be crucial. 
One has observed
${\cal B}(\Lambda_b\to \Lambda_c^+ D_s^-)\simeq 10^{-2}$ and
${\cal B}(\Sigma_c^{0(++)}\to \Lambda_c^+\pi^{-(+)})\simeq 100\%$~\cite{pdg},
indicating the sizeable weak and strong couplings of the baryon decays.
In addition,
the coupling of $X_{0(1)}^\prime\to D_s^-\pi$ is not small,
due to the $SU(3)$ flavor ($SU(3)_f$) symmetry
that has been enabled to relate $X_{0,1}^\prime$ and $X_{0,1}^0$ decays~\cite{He:2020jna}.
Hence,
${\cal B}(\Lambda_b\to \Sigma_c^{0(++)} X_{0,1}^{\prime\, 0(--)})$ are anticipated
to be as accessible as ${\cal B}(B^+\to D^+X_{0,1}^0)$ to the LHCb experiment.
In this paper, 
$\Lambda_b\to \Sigma_c^{0(++)} X_{0,1}^{\prime\, 0(--)}$
will be demonstrated as the promising decay channels
to confirm the existence of the $X_{0,1}^0$-like exotic states.

\section{Formalism}
\begin{figure}[t!]
\includegraphics[width=3.2in]{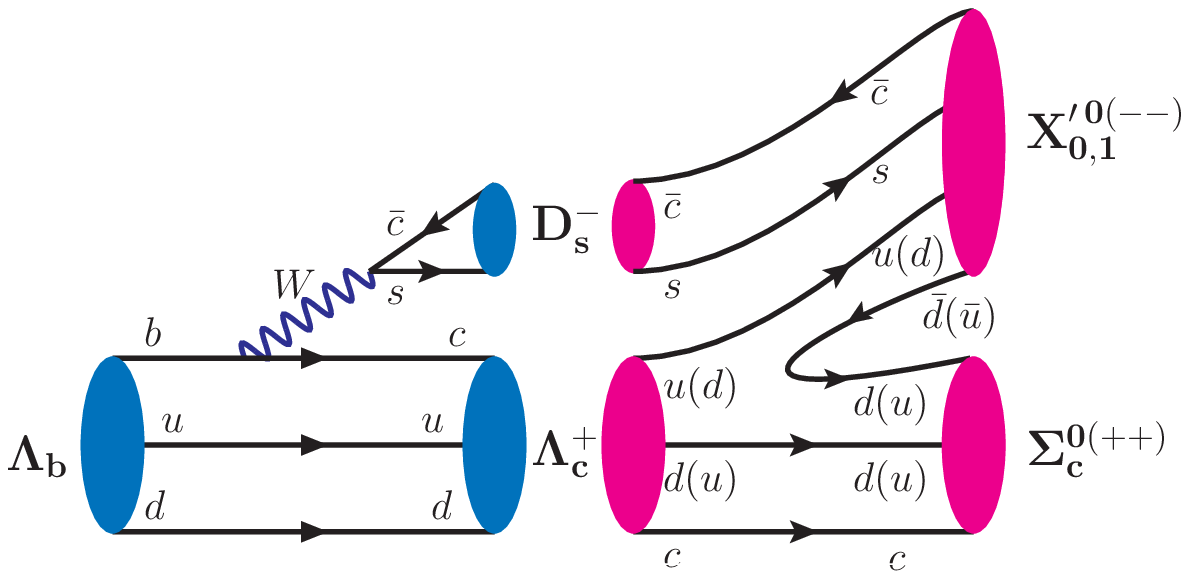}
\includegraphics[width=3.2in]{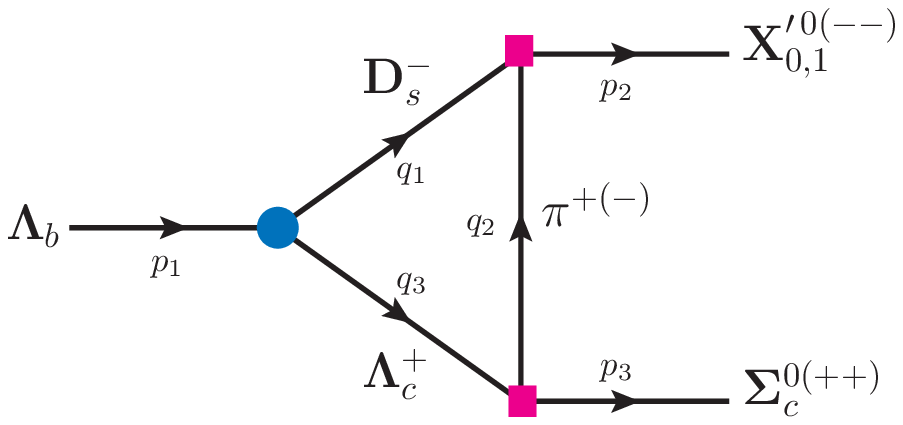}
\caption{Rescattering $\Lambda_{b}\to\Sigma_{c}X_{0,1}^{\prime}$ decays:
the left panel presents the quark lines,
and the right panel the momentum flows.}\label{triangle}
\end{figure}
The open-charm exotic states $X_{0,1}^0$ observed in $B^+\to D^+D^-K^+$
need to be confirmed by different decays.
We think $\Lambda_b\to \Sigma_c^{0(++)} X_{0,1}^{\prime\, 0(--)}$
may be promising. Like $X_{0,1}^0(\bar c d \bar s u)$,
$X_{0,1}^{\prime\,0}(\bar c s \bar d u)$
and $X_{0,1}^{\prime\,--}(\bar c s\bar u d)$
also consist of four different quark flavors.
See Fig.~\ref{triangle}, $\Lambda_b\to \Sigma_c X_{0,1}^{\prime}$
are the triangle-rescattering decays, separated into two parts.
The first part is the  weak decay $\Lambda_b\to  \Lambda_c^+ D_s^-$,
which proceeds through the $\Lambda_b\to  \Lambda_c^+$ transition
and $D_s^-$ meson production.
By following Refs.~\cite{Hsiao:2018zqd,Hsiao:2017umx,Leibovich:2003tw},
we derive the amplitude of $\Lambda_b\to  \Lambda_c^+ D_s^-$ as
\begin{eqnarray}\label{weakamp1}
{\cal M}_b(\Lambda_b\to  \Lambda_c^+ D_s^-)
&=&\bar u_{\Lambda_c}(F_b^+- F_b^-\gamma_5)u_{\Lambda_b}\,,
\end{eqnarray}
where
\begin{eqnarray}\label{weakamp2}
F_b^+=C_{\text w} m_-\bigg[f_1+\left(\frac{m_{D_s}^2}{m_+ m_-}\right)f_3\bigg]\,,\,\,\,\,\,\,\,\,\,
F_b^-= C_{\text w} m_+\bigg[g_1+\left(\frac{m_{D_s}^2}{m_+^2}\right)g_3\bigg]\,.
\end{eqnarray}
In the above, we define $m_\pm=m_{\Lambda_b}\pm m_{\Lambda_c}$ and
$C_{\text w}=i({G_F}/{\sqrt{2}})\,a_1V_{cb}V^*_{cs} f_{D_s}$, where
$G_F$, $V_{ij}$, $f_{D_s}$ and $f_{1,3}(g_{1,3})$ are the Fermi constant, CKM matrix element,
decay constant, and $\Lambda_b\to \Lambda_c^+$ transition form factors~\cite{Feldmann:2011xf},
respectively, while $a_1$ results from the factorization~\cite{Hsiao:2020gtc,Hsiao:2014mua}.
The second part as the rescattering effect proceeds through the strong decays,
such that $\Lambda_c^+ D_s^-$ are turned into $\Sigma_c X_{0,1}^{\prime}$
by emitting or receiving a pion. Accordingly,
the amplitudes are given by~\cite{He:2020jna,Albertus:2005zy,Can:2016ksz}
\begin{eqnarray}\label{strong}
{\cal M}_0(X_0^{\prime}\to D^{-}_{s}\pi)&=&g_0\,,\;
\hat{\cal M}_1(X_1^{\prime}\to D^{-}_{s}\pi)=g_1\epsilon\cdot(p_{D_s}-p_\pi)\,,\;\nonumber\\
{\cal M}_c(\Sigma_{c}\to \Lambda^{+}_{c}\pi)
&=&g_c \bar{u}_{\Lambda_{c}}\gamma_{5} u_{\Sigma_{c}}\,,
\end{eqnarray}
with $g_{c,0,1}$ the strong coupling constants and
$\epsilon_\mu$ the polarization four vector,
where ${\cal M}_c(\Sigma_{c}\to \Lambda^{+}_{c}\pi)$ 
is presented as that used in the quark model and lattice QCD.

To proceed, we assemble the two parts as the rescattering amplitudes,
given by~\cite{Yu:2020vlt,Hsiao:2019ait}
\begin{eqnarray}\label{Are}
{\cal M}(\Lambda_{b}\to \Sigma_{c} X_{0,1}^{\prime})&=&\int \frac{d^4{q}_{3}}{(2\pi)^{4}}
\frac{{\cal M}_b {\cal M}^{\dag}_c {\cal M}_{0,1} }
{(q_{1}^{2}-m_{D_{s}}^{2})(q_{2}^{2}-m_{\pi}^{2})(q_{3}^{2}-m_{\Lambda_{c}}^{2})}\,,
\end{eqnarray}
where ${\cal M}_1\equiv \hat{\cal M}_1 F_{\Lambda}(q_{2}^2)$,
and
$F_{\Lambda}(q_{2}^2)\equiv
(\Lambda^{2}-m^{2}_{\pi})/(\Lambda^{2}-q^{2}_{2})$
is the form factor that
takes care of the divergence in $\Lambda_{b}\to \Sigma_{c} X_1^{\prime}$.
Besides, $q_1=p_1-q_3$ and $q_2=p_3-q_3$
correspond to  the momentum flows in Fig.~\ref{triangle}.
In the general form, one expresses
the amplitudes of $\Lambda_b\to \Sigma_c X_{0,1}^{\prime}$ as
\begin{eqnarray}
{\cal M}(\Lambda_b\to \Sigma_c X_0^{\prime})&=&
\bar{u}_{\Sigma_{c}}(F_0^+-F_0^-\gamma_{5})u_{\Lambda_{b}}\,,\nonumber\\
{\cal M}(\Lambda_{b}\to \Sigma_{c} X_{1}^{\prime})&=&
\bar{u}_{\Sigma_{c}}[(F_1^+\gamma^{\mu}-F_1^-\gamma^{\mu}\gamma^{5})+
(G_1^+p_{3}^{\mu}-G_1^- p_{3}^{\mu}\gamma^{5})]u_{\Lambda_{b}}\epsilon_{\mu}\,.
\end{eqnarray}
To obtain $F_{0,1}^\pm$ and $G_1^\pm$, we need to
integrate over the variables  of the triangle loop in Eq.~(\ref{Are}),
for which the equations are given by~\cite{tHooft:1978jhc,Hahn:1998yk,Denner:2005nn}
\begin{eqnarray}\label{int}
\{C_{0};C^{\mu};C^{\mu\nu}\}&=&
\int \frac{d^{4}q_3}{i \pi^2}\frac{\{1;q_3^{\mu};q_3^{\mu}q_3^{\nu}\}}
{(q_3^{2}-m_{\Lambda_c}^{2})[(q_3-p_1)^{2}-m_{D_s}^{2}][(q_3-p_3)^{2}-m_{\pi}^{2}]}\,,
\end{eqnarray}
such that we obtain
\begin{eqnarray}\label{cij_1}
C^{\mu}&=&-p_1^\mu C_1 - p_3^\mu C_2\,,\nonumber\\
C^{\mu\nu}
&=&g^{\mu\nu}C_{00}+p_1^\mu p_1^\nu C_{11}
+p_3^\mu p_3^\nu C_{22}
+(p_1^\mu p_3^\nu+p_1^\nu p_3^\mu)C_{12}\,,
\end{eqnarray}
with the parameters $C_{0},C_1,C_2,C_{00},C_{11},C_{12},C_{22}$
to be given in the numerical analysis.
By replacing $m_{\pi}$ in Eq.~(\ref{int}) with $\Lambda$,
we obtain $\{C'_{0};C'^{\mu};C'^{\mu\nu}\}$ and then
define another set of parameters:
$\tilde{C}_{i(ij)}=C_{i(ij)}-C'_{i(ij)}$, where $i=(0,1,2)$ and $ij=(00,11,22,12)$.
We hence derive that
\begin{eqnarray}\label{cij_2}
F_0^\pm&=& \frac{\mp i g_c g_0 F_b^\mp}{16\pi^{2}}
\bigg(m_{\Lambda_{c}}C_0\pm m_{\Lambda_{b}}C_1+m_{\Sigma_{c}}C_2\bigg)\,,\;
F_1^\pm=\frac{ig_c g_1 F_b^\mp}{8\pi^2} \tilde{C}_{00}\,,\;\nonumber\\
G_1^\pm&=&\frac{-ig_c g_1 F_b^\mp}{8\pi^2}
\bigg[m_{\Lambda_{c}}(\tilde{C}_{0}+\tilde{C}_{1}+\tilde{C}_{2})\nonumber\\
&&
\pm m_{\Lambda_{b}}(\tilde{C}_{1}+\tilde{C}_{11}+\tilde{C}_{12})
+m_{\Sigma_{c}}(\tilde{C}_{2}+\tilde{C}_{12}+\tilde{C}_{22})\bigg]\,.
\end{eqnarray}

\section{Numerical results and Discussions}
In the numerical analysis, we adopt
$(V_{cb},V_{cs})=(A\lambda^2,1-\lambda^2/2)$
and $f_{D_s}=(249.9\pm 0.5)$~MeV, with
$A= 0.790\pm 0.017$ and $\lambda=0.22650\pm 0.00048$~\cite{pdg}.
We get $(f_1,g_1)=(0.59,0.53)$ and $(f_3,g_3)=(-0.02,-0.03)$
from the lattice QCD calculation~\cite{Detmold:2015aaa}.
Using Eqs.~(\ref{weakamp1}) and (\ref{weakamp2}), and
${\cal B}(\Lambda_b\to\Lambda_c^+ D_s^-)=(1.10\pm 0.10)\%$~\cite{pdg},
we extract $a_1=0.93\pm 0.04$, where $a_1$ of ${\cal O}(1.0)$ 
demonstrates the feasibility of the generalized factorization~\cite{Hsiao:2018zqd,
Hsiao:2017umx,Leibovich:2003tw}.
From $B(\Sigma_{c} \to \Lambda_{c}\pi)\simeq100\%$~\cite{pdg},
we determine $g_c=19.1$~GeV.
The $SU(3)_f$ symmetry has been enabled to relate
$X_{0,1}^{\prime\,0(--)}\to D_s^-\pi^{+(-)}$ and $X_{0,1}^0\to D^-K^+$,
such that we obtain $g_0=(2.85\pm0.32)$~GeV and $g_1=4.63\pm0.25$
from $\Gamma_{X_{0,1}^0}$ in Eq.~(\ref{MandW}); besides,
we take $m_{X_{0,1}^\prime}\simeq m_{X_{0,1}^0}$.

The integration of the triangle loop
in $\Lambda_b\to\Sigma_c X_0^\prime$ gives
\begin{eqnarray}
&&(C_0,C_1,C_2)=(0.37-0.32i,-1.50-2.30i,-0.91-0.55i)~\text{GeV}^{-2}\,.
\end{eqnarray}
On the other hand,
$\Lambda_b\to\Sigma_c X_1^\prime$ encounters the logarithmic divergence,
for which the cutoff $\Lambda$ needs to be introduced.
Since $\Lambda$ of ${\cal O}(1.0~\text{GeV})$ has been commonly used
as a phenomenological parameter~\cite{Tornqvist:1993ng,Li:1996yn,Wu:2019vbk},
we take $\Lambda=(1.00,1.25,1.50)$~GeV for the demonstration,
which results in\\
\begin{eqnarray}\label{CiCij}
\tilde{C}_0&=&(0.26+0.17i,0.36+0.10i,0.41+0.05i)~\text{GeV}^{-2}\,,\nonumber\\
\tilde{C}_1&=&(0.95-2.15i,0.35-2.75i,-0.11-2.94i)~\text{GeV}^{-2}\,,\nonumber\\
\tilde{C}_2&=&(0.06-0.24i,0.08-0.35i,0.05-0.45i)~\text{GeV}^{-2}\,,\nonumber\\
\tilde{C}_{00}&=&(-1.02-0.51i,-1.62-0.45i,-2.17-0.25i)\,,\nonumber\\
\tilde{C}_{11}&=&(-0.52+1.85i,-0.01+2.19i,0.33+2.23i)~\text{GeV}^{-2}\,,\nonumber\\
\tilde{C}_{12}&=&(-0.02+0.19i,0.00+0.26i,0.03+0.31i)~\text{GeV}^{-2}\,,\nonumber\\
\tilde{C}_{22}&=&(0.01+0.06i,0.02+0.08i,0.02+0.11i)~\text{GeV}^{-2}\,.
\end{eqnarray}

As a consequence, we obtain
\begin{eqnarray}\label{pre_br}
{\cal B}_0(\Lambda_{b}\to \Sigma_{c}^{0(++)} X_{0}^{\prime\,0(--)})
&=&(2.3\pm 0.6)\times 10^{-4}\,,\nonumber\\
{\cal B}_1(\Lambda_{b}\to \Sigma_{c}^{0(++)} X_{1}^{\prime\,0(--)})
&=&(4.3\pm 0.8^{+3.3}_{-2.5})\times 10^{-4}\,,
\end{eqnarray}
where ${\cal B}_{0,1}$ as large as $10^{-4}$
are not caused by the triangle singularity~\cite{Liu:2020orv}.
Indeed, the sizeable weak and strong coupling constants play the key role.
However, the calculations are at most as accurate as
the order of magnitude estimations due to the large uncertainties in Eq.~(\ref{pre_br}).
Explicitly, the first uncertainties combine
the errors from $V_{cb(cs)}$, $f_{D_s}$, $g_{0,1}$, $a_1$, and
the second ones from $\Lambda$,
where we simply take $\Lambda=(1.25\pm 0.25)$~GeV for a naive estimate
of ${\cal B}(\Lambda_{b}\to \Sigma_{c} X_{1}^{\prime})$,
in accordance with $\tilde C_i$ and $\tilde C_{ij}$ in Eq.~(\ref{CiCij}).

In the heavy-baryon chiral perturbation theory~\cite{Cheng:2015naa,Yan:1992gz},
the amplitude of $\Sigma_{c}\to \Lambda^{+}_{c}\pi$ in Eq.~(\ref{strong})
has another form: ${\cal M}'_c(\Sigma_{c}\to \Lambda^{+}_{c}\pi)=
g'_c\bar u_{\Lambda_c}(\gamma_\mu\gamma_5 p_\pi^\mu) u_{\Sigma_c}$,
where $g'_c=4.0$. Subsequently, $p_\pi^\mu$ added to the triangle loop
causes the logarithmic (linear) divergence for 
$\Lambda_{b}\to \Sigma_{c}^{0(++)} X_{0(1)}^{\prime\,0(--)}$, 
which corresponds to the integration with $C^{\mu\nu}$ ($C^{\mu\nu\rho}$). 
Note that $C^{\mu\nu\rho}$ has an extra $q_3^\rho$ 
compared to $C^{\mu\nu}$ in Eq.~(\ref{int}),
whose detailed form can be found in~\cite{tHooft:1978jhc,Hahn:1998yk,Denner:2005nn}.
We hence obtain different results, given by
\begin{eqnarray}\label{pre_br2}
{\cal B}^\prime_0(\Lambda_{b}\to \Sigma_{c}^{0(++)} X_{0}^{\prime\,0(--)})
&=&(3.1\pm 0.9^{+0.5}_{-0.9})\times 10^{-4}\,,\nonumber\\
{\cal B}^\prime_1(\Lambda_{b}\to \Sigma_{c}^{0(++)} X_{1}^{\prime\,0(--)})
&=&(4.5\pm 0.8^{+3.1}_{-2.5})\times 10^{-3}\,,
\end{eqnarray}
where ${\cal B}^\prime_0(\Lambda_{b}\to \Sigma_{c}X_{0})$ 
is slightly deviated from ${\cal B}_0(\Lambda_{b}\to \Sigma_{c}X_{0})$;
however, due to the cutoff, it is more uncertain.
On the other hand, ${\cal B}^\prime_1(\Lambda_{b}\to \Sigma_{c}X_{1})$ 
is 10 times larger than ${\cal B}_1(\Lambda_{b}\to \Sigma_{c}X_{1})$,
presenting a sensitivity to $p_\pi^\mu$ from the linear divergence.
To distinguish between the two different strong couplings,
${\cal B}_0/{\cal B}_1\simeq 0.5$ and ${\cal B}'_0/{\cal B}'_1\simeq 0.07$
can be used for the future experimental examination.

The exotic $X_{0,1}^0$ are observed in $B^+\to D^+ D^- K^+$.
Likewise, we expect $X_0^{\prime\,0(--)}$ to be observed
in $\Lambda_b\to \Sigma_c^{0(++)} M_c M$,
which receives the resonant contributions from
$\Lambda_b\to \Sigma_c^{0(++)}X_0^{\prime\,0(--)},X_0^{\prime\,0(--)}\to M_c M$
with $M_c M=D_s^-\pi^+$, $\bar D^0 \bar K^0$ $(D_s^-\pi^-$, $D^-K^-)$.
Approximately, we present the resonant branching fractions as
\begin{eqnarray}
&&{\cal B}(\Lambda_b\to \Sigma_cX_{0,1}^{\prime}, X_{0,1}^{\prime}\to M_c M)
\simeq
{\cal B}(\Lambda_b\to \Sigma_c X_{0,1}^{\prime}){\cal B}(X_{0,1}^{\prime}\to M_cM)\,,
\end{eqnarray}
with ${\cal B}(\Lambda_b\to \Sigma_c X_{0,1}^{\prime})$ in Eq.~(\ref{pre_br}).
In the $SU(3)_f$ symmetry,
${\cal B}(X_{0,1}^{\prime}\to M_cM)$
are given by~\cite{He:2020jna}
\begin{eqnarray}
&&{\cal B}(X_{0,1}^{\prime\,0(--)}\to D_s^-\pi^\pm)
=(51,53)\%\,,\nonumber\\
&&{\cal B}(X_{0,1}^{\prime\,0(--)}\to \bar D^0 \bar K^0(D^- K^-))
=(49,47)\%\,.
\end{eqnarray}
We can hence estimate 
${\cal B}(\Lambda_{b}\to \Sigma_{c}^{++}X_{0}^{\prime\,--}
X_{0,1}^{\prime\,--}\to D^-K^-)$ as large as $10^{-4}$, 
which seems to be very accessible to the LHCb experiment. However,
there might exist other resonant decays to interfere with
$\Lambda_{b}\to \Sigma_{c}^{++}X_{0}^{\prime\,--},X_{0,1}^{\prime\,--}\to D^-K^-$, 
which would cause a complicated amplitude analysis. For example,
the amplitude analysis of $B^+\to D^+ X_{0,1}^0,X_{0,1}^0\to D^- K^+$ is complicated,
which is due to the resonant decays $B^+\to M_{c\bar c} K^+,M_{c\bar c}\to D^+ D^-$
with $M_{c\bar c}=\psi(3770)$, $\chi_{c0(c2)}(3770)$, and $\psi(4040,4160,4415)$
to interfere with
$B^+\to D^+ X_{0,1}^0,X_{0,1}^0\to D^- K^+$~\cite{Aaij:2020hon,Aaij:2020ypa}.

Apart from
$\Lambda_b\to\Lambda_c^+ D_s^-\to\Sigma_c X_{0,1}^\prime$ with $\pi$ exchange,
other $b$-baryon decays can also produce the $X_{0,1}$-like states
of $Y_{1}^{\prime\,-}(\bar c s \bar n n)$, $Y_{1}^-(\bar c n \bar s n)$, and
$\bar X_{0,1}^{\prime\,0}(c\bar s d\bar u)$ in Eq.~(\ref{su3}).
Taking the quark level $b\to c\bar c s$, $c\bar cd$ and $c \bar u d$ weak decays
as the examples, the possible rescattering decays are given by
\begin{eqnarray}
&&b\to c\bar{c}s:\;\nonumber\\
&&
\Lambda_b\to\Lambda_c^+ D_s^-\to\Sigma^{+}_c Y_{1}^{\prime\,-}\,,\;
\Xi_{b}^{0(-)}\to\Xi_{c}^{+(0)}D_s^-\to\Xi_{c}^{+(0)}(2645)Y_{1}^{\prime\,-}\;\;
\;\text{(with $\pi^{0}$ exchange)}\,,\nonumber\\
&&
\Xi_{b}^{0}\to\Xi_{c}^{+}D_s^-\to\Xi_{c}^{0}(2645)X_{0,1}^{\prime\,0}\,,\;
\Xi_{b}^{-}\to\Xi_{c}^{0}D_s^-\to\Xi_{c}^{+}(2645)X_{0,1}^{\prime\,--}\;\;
\text{(with $\pi^{\pm}$ exchange)}\,,\nonumber\\
&&b\to c\bar{c}d:\;\nonumber\\
&&
\Lambda_{b}\to\Lambda_{c}^{+}D^-\to\Xi_{c}^{\prime\,0}  X_{0,1}^{0}
\;\;\;\;\text{(with $K^\pm$ exchange)}\,,\nonumber\\
&&b\to c\bar{u}d:\;\nonumber\\
&&
\Lambda_{b}\to\Lambda_{c}^{+}\pi^-\to\Lambda^{(*)} \bar X_{0,1}^{\prime\,0}\,,\;
\Lambda_{b}\to\Lambda_{c}^{+}\pi^-\to\Sigma^{(*)0} \bar X_{0,1}^{\prime\,0}\,,\nonumber\\
&&
\Xi_{b}^{0(-)}\to\Xi_{c}^{+(0)}\pi^-\to\Xi^{(*)0(-)} \bar X_{0,1}^{\prime\,0}\;\;
\;\;\;\;\text{(with $D_s^\pm$ exchange)}\,,
\end{eqnarray}
where ${\bf B}^*$ stands for the higher-wave baryon state,
and $\Xi_{c}(2645)$ is able to decay into $\Xi_c\pi$.
These decays are worthy of the future explorations;
particularly, we estimate that
${\cal B}(\Lambda_b\to\Sigma^{+}_c Y_{1}^{\prime\,-})\sim 10^{-4}$,
${\cal B}(\Xi_{b}^{0(-)}\to \Xi_{c}^{0(+)}(2645)X_{0}^{\prime\,0(--)})\sim 10^{-5}$ and
${\cal B}(\Lambda_{b}\to \Xi_{c}^{\prime\,0}X_{0,1}^{0})\sim 10^{-5}$,
which suggest interesting measurements.
As the final remark, we emphasize that our estimations
rely on the observed weak decays in the triangle loop,
such as the color-allowed processes
$\Lambda_b\to \Lambda_c^+ D_s^-$, $\Xi_{b}^{0(-)}\to\Xi_{c}^{+(0)}D_s^-$ and
$\Lambda_{b}\to\Lambda_{c}^{+}D^-(\pi^-)$. However,
there should be other triangle rescatterings
to be discovered in the future work. For example,
those proceed through the similar color-allowed
$\Lambda_b$ decay processes but with $\Lambda_c^+ D_s^-$ or $\Lambda_{c}^{+}D^-(\pi^-)$ 
replaced by other particles of the same quark contents,
or those through the color-suppressed ones,
which are not necessarily suppressed in the final state interaction.
Moreover, when the open-charm exotic particles are interpreted
as the bound states~\cite{Molina:2010tx},
it allows for the rescatterings with the exchanges of the vector mesons,
which also deserve future investigation.

\section{Conclusions}
In summary, we have proposed the rescattering decays
$\Lambda_b\to\Lambda_c^+ D_s^-\to\Sigma_c X_{0,1}^\prime$ with the $\pi$ exchange,
in order to provide the different decays
to confirm the $X_{0,1}$-like exotic states observed in $B^+\to D^+D^-K^+$.
As the order of magnitude estimates, we have calculated that
${\cal B}(\Lambda_b\to \Sigma_c^{(++)} X_0^{\prime\,0(--)})
=(2.3\pm 0.6)\times 10^{-4}$ and
${\cal B}(\Lambda_b\to \Sigma_c^{0(++)} X_1^{\prime\,0(--)})
=(4.3\pm 0.8^{+3.3}_{-2.5})\times 10^{-4}$.
Other possible $b$-baryon decays with the $X_{0,1}$-like exotic states
have also been estimated, such as
${\cal B}(\Lambda_b\to\Sigma^{+}_c Y_{1}^{\prime\,-})\sim 10^{-4}$,
${\cal B}(\Xi_{b}^{0(-)}\to \Xi_{c}^{0(+)}(2645)X_{0}^{\prime\,0(--)})\sim 10^{-5}$ and
${\cal B}(\Lambda_{b}\to \Xi_{c}^{\prime\,0}X_{0,1}^{0})\sim 10^{-5}$.
To measure $X_{0,1}^{\prime\,0(--)}\to M_c M$
in the $M_c M$ invariant mass spectrum, where
$M_c M$ can be $D_s^-\pi^+$, $\bar D^0 \bar K^0$ $(D_s^-\pi^-$, $D^-K^-)$,
we have introduced
the three-body decay channel $\Lambda_b\to \Sigma_c^{0(++)} M_c M$.
The branching fraction estimated at the level of $10^{-4}$
can be accessible to the near future measurements.

\section*{ACKNOWLEDGMENTS}
YY would like to thank Dr.~Xu-Chang Zheng  for useful discussions.
YKH was supported in part by NSFC (Grant No.~11675030).
YY was supported in part by NSFC (Grant No.~11905023) and
CQCSTC (Grants No. cstc2020jcyj-msxmX0555, No. cstc2020jcyj-msxmX0810).

\end{document}